# Position-enhanced and Time-aware Graph Convolutional Network for Sequential Recommendations

Liwei Huang[1,2], Yutao Ma[3], Yanbo Liu[2], Bohong (Danny) Du[4], Shuliang Wang[1], and Deyi Li[5]

1. School of Computer Science and Technology, Beijing Institute of Technology, Beijing 100081, China
2. Beijing Institute of Remote Sensing, Beijing 100854, China
3. School of Computer Science, Wuhan University, Wuhan 430072, China
4. Department of Computer Science, Stanford University, Stanford, CA 94305, USA
5. Chinese Academy of Engineering, Beijing, 100088, China

E-mails: dr_huanglw@163.com, ytma@whu.edu.cn, liuyanbonudt@163.com, dannydu@stanford.edu, slwang2011@bit.edu.cn, lidy@cae.cn

**Abstract:** The sequential recommendation (also known as the next-item recommendation), which aims to predict the following item to recommend in a session according to users' historical behavior, plays a critical role in improving session-based recommender systems. Most of the existing deep learning-based approaches utilize the recurrent neural network architecture or self-attention to model the sequential patterns and temporal influence among a user's historical behavior and learn the user's preference at a specific time. However, these methods have two main drawbacks. First, they focus on modeling users' dynamic states from a user-centric perspective and always neglect the dynamics of items over time. Second, most of them deal with only the first-order user-item interactions and do not consider the high-order connectivity between users and items, which has recently been proved helpful for the sequential recommendation. To address the above problems, in this article, we attempt to model user-item interactions by a bipartite graph structure and propose a new recommendation approach based on a **P**osition-enhanced and **T**ime-aware **G**raph **C**onvolutional **N**etwork (PTGCN) for the sequential recommendation. PTGCN models the sequential patterns and temporal dynamics between user-item interactions by defining a position-enhanced and time-aware graph convolution operation and learning the dynamic representations of users and items simultaneously on the bipartite graph with a self-attention aggregator. Also, it realizes the high-order connectivity between users and items by stacking multi-layer graph convolutions. To demonstrate the effectiveness of PTGCN, we carried out a comprehensive evaluation of PTGCN on three real-world datasets of different sizes compared with a few competitive baselines. Experimental results indicate that PTGCN outperforms several state-of-the-art sequential recommendation models in terms of two commonly-used evaluation metrics for ranking. In particular, it can make a better trade-off between recommendation performance and model training efficiency, which holds great potential for online session-based recommendation scenarios in the future.

# 1 Introduction

## 1.1 Background

The purpose of recommender systems is to recommend relevant items to users. Until now, they have achieved great success in many web applications. Massive user behavior records collected from web applications provide an unprecedented opportunity for recommender systems to achieve accurate recommendations. The chronology of online user behavior is vital to capture sequential patterns for developing better recommender systems. As an emerging recommendation scenario, the sequential recommendation (or the next-item recommendation), which aims to predict users' future behaviors based on historical action sequences, has attracted increasing attention in academic and industrial fields.

Most of the previous sequential recommendation algorithms (or models) focus on the one-directional chain structure of action sequences sorted by interaction time, including two approaches: the Markov chain-based approach and the neural network-based approach. The Markov chain-based approach [1], [2], [3] makes recommendations based on the recent $L$ actions using an $L$-order Markov chain. By simplifying some assumptions, this approach can achieve good results in high-sparsity settings. However, it often underperforms in long-term recommendation scenarios due to the limited ability to model the intricate dynamics of user-item interactions [4], [5]. Compared with the Markov chain-based approach, the recommendation approach based on neural networks, such as recurrent neural networks (RNNs) [6], [7], convolutional neural networks (CNNs) [8], and Transformer [5], [9], has become very popular to model sequential patterns in online user behavior. Recently, some neural network-based methods [9], [10], [11] attempted to utilize the temporal dynamics of user behavior to improve recommendation performance in specific domains.

## 1.2 Motivation

Although these RNN- and Transformer-based methods can obtain good results in the sequential recommendation task, they still have two shortcomings. First, most of them [5], [6], [7], [8], [9], [12] consider only the temporal dynamics of user behavior while neglecting the temporal dynamics of item properties. As we know, items have static properties that do not change over time and time-evolving properties. An item may show different temporal dynamics over time, such as growth in popularity and social topic drift [13]. It is necessary to design a unified framework to simultaneously leverage the dynamics of both user behavior and item properties. Second, most existing methods only consider direct user-item interactions (i.e., the first-order connectivity) in defining the loss function for model training and neglect important collaborative information embedded in user-user and item-item interactions. As a result, the embeddings of users and items may be insufficient to capture the collaborative signal (or called the high-order connectivity) [14], which represents the behavioral similarity between users (or items).

Fig. 1 illustrates the high-order connectivity in a bipartite graph derived from historical user-item interactions. The bipartite graph in Fig. 1(a) includes three users ($u_1, u_2, u_3$), four items ($v_1, v_2, v_3, v_4$), and the interactions between them, each of which has the timestamp ($t$) the interaction occurred. Fig. 1(b) shows two tree-like structures rooted in item $v_2$, which denote the high-order connectivity of $v_2$ at $t_5$ and $t_4$, respectively. Obviously, $v_2$ has different connectivities at the two moments ($t_5$ and $t_4$) because $v_2$ has a new interaction with user $u_1$ at $t_5$. Therefore, it is valuable to consider the temporal dynamics of items in the sequential recommendation. Besides, the high-order connectivity contains rich semantics that carries the collaborative signal. For example, there are two paths between $v_2$ and $v_4$ at $t_5$ (i.e., $v_2 \to u_1 \to v_4$ and $v_2 \to u_3 \to v_4$), suggesting that there is a high similarity between $v_2$ and $v_4$ at $t_5$. Considering that $u_2$ interacted with $v_4$ at $t_2$, it is more likely to recommend $v_2$ to $u_2$ at $t_5$. If we consider only the first-order connectivity, we cannot make an appropriate recommendation in terms of the similarity between $v_2$ and $v_4$ at $t_5$. Hence, it is essential to model the high-order connectivity on bipartite graphs to characterize user preference better.

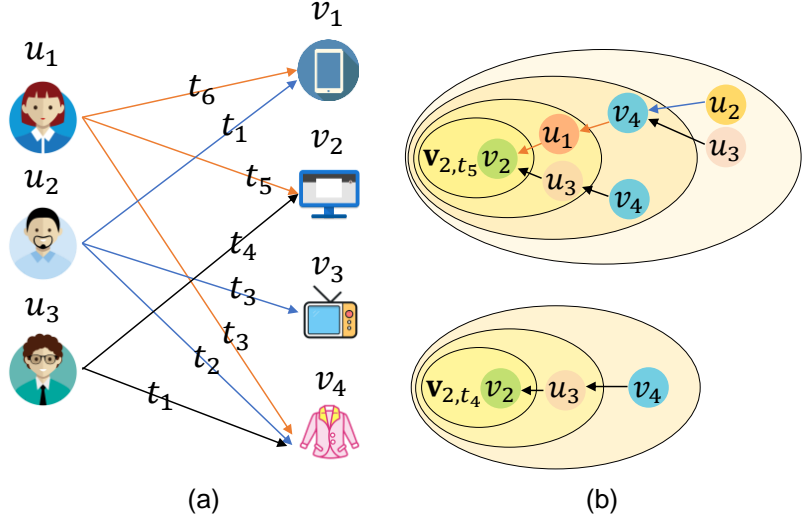

Fig. 1. An example of the high-order connectivity in a user-item interaction graph.

Due to the powerful modeling capability of graph neural networks (GNNs) on different graphs, they have also been applied to general-purpose recommender systems in recent years. Subsection 2.1 summarizes the related work based on GNNs, especially on graph conventional networks (GCNs). However, it is hard for these GCN-based recommendation models that failed to capture short- and long-term sequential information to work well in the sequential recommendation scenario [15], [16]. Since action sequences are not inherently graphs for GNN training [17], some recent researches attempted to address such a problem with different sequence-to-graph solutions. One straightforward solution to modeling user preference is applying the GNN architecture and attention mechanism to a graph built from item sequences [16], [17], [18]. An alternative solution is encoding the high-order connectivity with an attentive GCN in the sequential context [19], which can better represent the behavioral similarity between users. However, none of these GCN-based sequential recommendation models can explicitly encode the high-order connectivity between users and items in a user-item interaction graph by simultaneously modeling the temporal dynamics of users and items in a unified manner. This challenging problem is worthy of investigation for sequential recommender systems.

### 1.3 Contribution

To address the challenge mentioned above, we propose a **P**osition-enhanced and **T**ime-aware **G**raph **C**onvolutional **N**etwork (PTGCN) model for the sequential recommendation task. More specifically, we construct a bipartite graph with the interactions between users and items and design a specific graph convolution with time and order information to express user (or item) embedding by integrating the hidden features of items (or users) with which the user (or item) has recently interacted. On the one hand, we propose a self-attention aggregator in the graph convolution operation, which can simultaneously model sequential patterns of users and items to learn the dynamic embeddings of users and items at each moment. On the other hand, we perform multi-layer graph convolutions on the user-item bipartite graph to learn the collaborative signal of each node and the relations between two consecutive orders of connectivity (e.g., the first- and second-order user-item interactions). In this way, we can structure a more expressive model to capture the high-order connectivity between users and items in the corresponding user-item interaction graph. Besides, the experimental results on three publicly available datasets indicate the advantages of PTGCN over some selected state-of-the-art (SOTA) methods in terms of two commonly-used evaluation metrics for ranking. In brief, the technical contributions of this study are summarized as follows.
- We propose a GCN-based model with self-attention for the sequential recommendation task. It can explicitly encode the high-order connectivity in the user-item bipartite graph to learn the dynamic representations of users and items via multi-layer graph convolutions in a unified manner, which has never been reported in previous studies on the sequential recommendation.
- We design a self-attention aggregator to adaptively obtain the dynamic representation of each user (or item) by integrating the hidden features of the items (or users) with which the user (or item) interacted recently. The aggregator can simultaneously model sequential patterns and temporal influence, which are of value for sequential recommendations.
- We conduct extensive experiments on three real-world datasets of different sizes. The experimental results demonstrate the advantages of PTGCN over a few competitive baselines regarding two commonly-used evaluation metrics. In particular, PTGCN can make a better trade-off between recommendation performance and model training efficiency.

### 1.4 Organization

The remainder of this article is organized as follows. Section 2 reviews the work related to the sequential recommendation in the field of recommender systems. The problem to resolve in this study is formulated in Section 3. Section 4 presents the proposed PTGCN model in detail. Experiment setups and results are presented in Section 5. Finally, Section 6 concludes this paper and provides an overview of our future work.

## 2 Related Work

### 2.1 Recommender Systems based on Graph Convolutional Networks

Along with the popularity of GNNs, researchers recently proposed a series of models for recommendation [14], [20], [21], [22], [23], [24], [25], [26] by combining GCNs with traditional recommendation techniques to take advantage of rich structural information embodied in the interactions between users and items. For example, Berg *et al*. [21] proposed a graph convolutional matrix completion (GC-MC) framework from the perspective of link prediction on graphs. However, this framework only models the direct ratings by users on items with one convolutional layer. Therefore, it cannot effectively capture the high-order collaborative information between users and items. Ying *et al*. [20] developed a GCN-based algorithm for Pinterest image recommendation (PinSage) that employs multiple graph convolutional layers on the item-item interaction graph. However, it only models the collaborative signal on the level of item relations. Zheng *et al*. [22] proposed a spectral collaborative filtering (SpectralCF) method that uses a spectral convolution operation to predict all possible connections between users and items to alleviate the cold-start problem. However, SpectralCF has high computational complexity and is unsuitable for scaling up to large-scale recommender systems.

Wang *et al*. [14] then designed a neural graph collaborative filtering (NGCF) model by propagating the embeddings of users and items on the user-item bipartite graph, which was proven to learn a more expressive collaborative signal for the target user. Zhao *et al*. [23] proposed a framework named IntentGC to leverage explicit preferences and heterogeneous relationships by a GCN. They also designed a faster graph convolutional model called IntentNet to help apply IntentGC to web-scale applications. Chen *et al*. [24] found that removing non-linearities in GCNs could improve recommendation performance. They, therefore, proposed a residual network structure that can alleviate the over-smoothing problem in the graph convolution aggregation operation with sparse user-item interactions. Sun *et al*. [25] proposed a neighbor interaction aware framework based on GCNs, which can explicitly model the relationships between neighbor nodes and exploit the heterogeneous nature of user-item bipartite graphs. He *et al*. [26] empirically found that feature transformation and nonlinear activation in GCNs contribute little to the recommendation performance of collaborative filtering. They then simplified the design of GCN components and proposed a new model named LightGCN, which includes only the neighborhood aggregation component in GCNs for collaborative filtering.

The methods mentioned above were initially designed for general-purpose recommendations, such as the item and social recommendations. In other words, they did not consider the chronological order of user behavior; also, they did not model the temporal information of the interactions between users and items. Hence, these GCN-based recommendation methods cannot be directly applied to the sequential recommendation task. Recently, a few researchers designed and implemented GCN-based models for specific graphs constructed on item sequences in the sequential recommendation scenario [16], [17], [18]. For example, Chang *et al.* [18] proposed a graph neural network that leverages graph convolutional propagation and graph pooling to extract implicit preference signals on an item-item interest graph. Similarly, Ma *et al.* [17] built an item graph by extracting each item and its three following items from an item sequence and adding edges between them; also, they modeled item-occurrence patterns. However, these models cannot model the high-order connectivity between users and items over homogeneous graphs; besides, they neglected the time-evolving properties of items. Unlike previous GCN-based sequential recommendation models, this study aims to effectively utilize the high-order collaborative signal in the user-item bipartite graph and model sequential patterns and the temporal dynamics of users and items.

**2.2 Sequential Recommendation**

Unlike the general-purpose recommendation, the sequential recommendation organizes all historical interactions in chronological order of events. It then predicts the items with which the target user may interact soon. A typical solution to this problem is the Markov chain-based approach. For example, Rendle *et al*. [1] modeled long-term user preferences and short-term transitions over items by factorizing personalized Markov chains. In [2], a translation-based method was proposed for the sequential recommendation to model three-order interactions between a user, the user's previously-visited item, and the following item to visit. Similarly, Pasricha *et al.* [27] proposed a hybrid approach that combines transition-based methods for sequential recommendation and factorization machines. By modeling pairwise user-item and item-item interactions, He *et al.* [3] integrated similarity-based models with high-order Markov chains to realize personalized sequential recommendations. However, the Markov chain-based approach and its improved versions often suffer from one of the apparent disadvantages of Markov chains, i.e., the limited ability to perform mid-term and long-term forecasts well.

Inspired by the tremendous progress of sequence learning in natural language processing (NLP), deep learning has been widely used to learn long-term user preferences and sequential patterns in recommender systems. For example, Hidasi *et al.* [6], [7] introduced the RNN architecture with novel ranking loss functions to the session-based recommendation. A few RNN variants, such as long and short-term memory (LSTM) [28] and gated recurrent unit (GRU), have also been proposed to enhance recommendation performance by leveraging attention mechanisms [11], [29], memory networks [4], [30], [31], the hierarchical structure [32], [33], and so on. For example, Wang *et al.* [31] proposed a collaborative session-based recommendation machine (CSRM), including two memory modules. CSMR utilizes a fusion gating mechanism to selectively integrate information from the two memory modules. Besides, the CNN architecture recognized in computer vision was applied to capture short-term context information in the sequential recommendation task [8]. The proposed method embeds an item sequence into an "image" and learns sequential patterns as local features of the image by the convolution operation. Huang *et al*. [34] then proposed a graph multi-scale pyramid network with convolutional-recurrent encoders to extract the categorical-temporal pattern of user behavior at each time scale. Also, they designed a resolution (scale)-wise recalibration gating mechanism that is essentially an MLP to weigh representations from different scales adaptively.

Recently, self-attention [35] has shown promising performance in different NLP tasks. It is therefore applied to the sequential recommendation. In [5], [9], [36], [37], [38], [39], [40], [41], [42], researchers utilized only self-attention to model sequential patterns of user behavior rather than using the RNN architecture. Instead, these methods achieved better performance and efficiency. For example, Zhou *et al*. [42] introduced a self-supervised learning method with mutual information maximization to the self-attentive network architecture to deal with insufficient training data. They also extended the proposed method to other recommendation models to improve recommendation performance.

Due to the success of GNNs in recommender systems, a specific kind of item-item interaction graph, hypergraph, was recently used to model many-to-many and high-order relations among items in the sequential recommendation task [43], [44]. Each hyperedge in a hypergraph is set-like and contains two or more nodes. Feng *et al*. [45] designed a hyperedge convolution operation, and Yadati *et al*. [46] proposed a novel GCN for semi-supervised learning on attributed hypergraphs. According to the above work, Wang *et al*. [44] attempted to model short-term user preference for next-item recommendation using hypergraph, and Xia *et al*. [43] proposed a dual-channel hypergraph convolutional network (DHCN) to improve session-based recommendation. Moreover, self-supervised learning was integrated into the training process of the DHCN network to enhance hypergraph modeling [43]. In addition to hypergraphs, researchers have proposed sequential recommendation models based on other types of graphs, such as tripartite graphs [19] and heterogeneous interaction graphs [47], to capture the complex and rich structural information of user-item interactions. However, these models built on the complex graph structure always have relatively low model training efficiency.

Overall, our work is different from the above studies in two aspects. First, the above studies focus on modeling user preference (or user behavior), while our work aims to model the temporal dynamics of users and items simultaneously. Second, we attempt to effectively model the high-order collaborative information between users and items using a new self-attentive GCN, which incorporates sequence position information and temporal information into the designed graph convolution operation.

## 3 Problem Definition

This section presents primary notations used in this article (see Table 1) and formulates the research problem in sequential recommendations.

*Definition 1 (Interaction)*. Let $U$ and $V$ denote the user and item set, respectively. An interaction $i_{u,v,t}$ is an action that occurs between user $u \in U$ and item $v \in V$ at time point $t$, represented with a quaternion $i_{u,v,t} = (u, v, p, t)$, where $p$ is the position index in the ordered set of interactions (sorted in chronological order).

*Definition 2 (User-Item Interaction Graph)*. A user-item interaction graph is a bipartite graph $G$,

whose vertex set can be decomposed into two disjoint sets $U$ and $V$. Each edge in the graph $G$, which links a vertex in $U$ to a vertex in $V$, represents an interaction.

*Definition 3 (User Neighborhood).* A user's neighborhood is an ordered subset of interactions performed by the user $u$, including the latest $n$ interactions before time point $t_q$, denoted as $N_{u,t_q} = \{i_{u,v,t_m} | v \in V, q - n < m \leq q\}$.

*Definition 4 (Item Neighborhood).* An item's neighborhood is an ordered subset of interactions with the item $v$, denoted as $N_{v,t_q} = \{i_{u,v,t_m} | u \in U, q - n < m \leq q\}$, which contains the latest $n$ interactions before $t_q$.

Many GCNs, such as GraphSAGE [48], obtain a fixed-size neighborhood through sampling. The time-aware user neighborhood (or item neighborhood) defined in this work is an ordered set composed of the $n$ latest interactions by the target user (or with the target item). If the number of historical interactions before the current time does not reach $n$, we will use a padding operation.

*Definition 5 (Node Flow).* A node's node flow is a tree-like structure rooted in the node. It consists of the root node and $M$ ($M \geq 1$) layers, each of which has a collection of nodes of the same type sampled via the neighborhood. More specifically, the target node's neighbors reached in different hops in a user-item interaction graph are placed in different layers in the node flow.

In this study, we use the concept of node flow to characterize the high-order connectivity between nodes in a bipartite graph. The high-order connectivity can be easily realized in practice by the breadth-first search algorithm or other improved algorithms. Then, we define the research problem in the sequential recommendation task as below.

*Definition 6 (Sequential recommendation).* Given a user-item interaction graph constructed from historical interactions $\{i_{u,v,t}\}$, for the target user $u$ at time point $t_N$, the goal of the sequential recommendation is to predict the most likely item $v$ with which user $u$ will interact at $t_{N+1}$.

Table 1. Primary notations.

| Symbol | Description |
|---|---|
| $U$, $V$ | the set of users and the set of items |
| $i_{u,v,t} = (u, v, p, t)$ | an interaction occurs between user $u$ and item $v$ at time point $t$ |
| $G = <(U,V), I>$ | a user-item interaction graph |
| $N_{u,t_q}$ | the user neighborhood of user $u$ at time point $t_q$ |
| $N_{v,t_q}$ | the item neighborhood of item $v$ at $t_q$ |
| $NF_{u,t}$ | the node flow of user $u$ at $t$ |
| $NF_{v,t}$ | the node flow of item $v$ at $t$ |
| $\mathbf{u}_{i,t}$ | the dynamic embedding of user $u_i$ at $t$ |
| $\mathbf{v}_{j,t}$ | the dynamic embedding of item $v_j$ at $t$ |
| $d$ | the latent dimension of user embeddings and item embeddings |
| $\mathbf{t}$ | the embedding of $t$ |
| $\mathbf{p}_{i,t}$, $\mathbf{p}_{j,t}$ | the position embeddings of $u_i$ and $v_j$ at $t$ |

# 4 Position-enhanced and Time-aware GCN

## 4.1 Overall Framework

PTGCN learns the dynamic embeddings of users and items at different moments to further predict possible interactions between the target user and items at the next moment. The main challenges that PTGCN has to face are two-fold: (1) model sequential patterns and temporal influence of interactions simultaneously; (2) capture the high-order collaborative information between users and items and update the user (or item) embedding timely. In this study, PTGCN defines a position-enhanced and time-aware graph convolution operation to solve the above two challenges. More specifically, for the first challenge,

PTGCN defines a self-attention aggregator in the graph convolution operation to model sequential patterns of user behavior and the temporal dynamics of interactions in a unified way. For the second challenge, PTGCN generates the dynamic embeddings of users and items using the graph convolution operation and then stacks multi-layer graph convolutions to model the high-order collaborative information.

Fig. 2 illustrates the architecture of PTGCN, which is a model framework with three components: (1) an embedding layer that generates four types of embeddings, namely user embedding, item embedding, time embedding, and position embedding; (2) the convolutional layer that refines the embeddings of users and items by modeling the high-order connectivity with the designed position-enhanced and time-aware graph convolution; (3) a prediction layer that aggregates the refined user embeddings and item embeddings and then outputs a score for each user-item pair. For more details of the three PTGCN components, please refer to Subsections 4.2, 4.3, and 4.4.

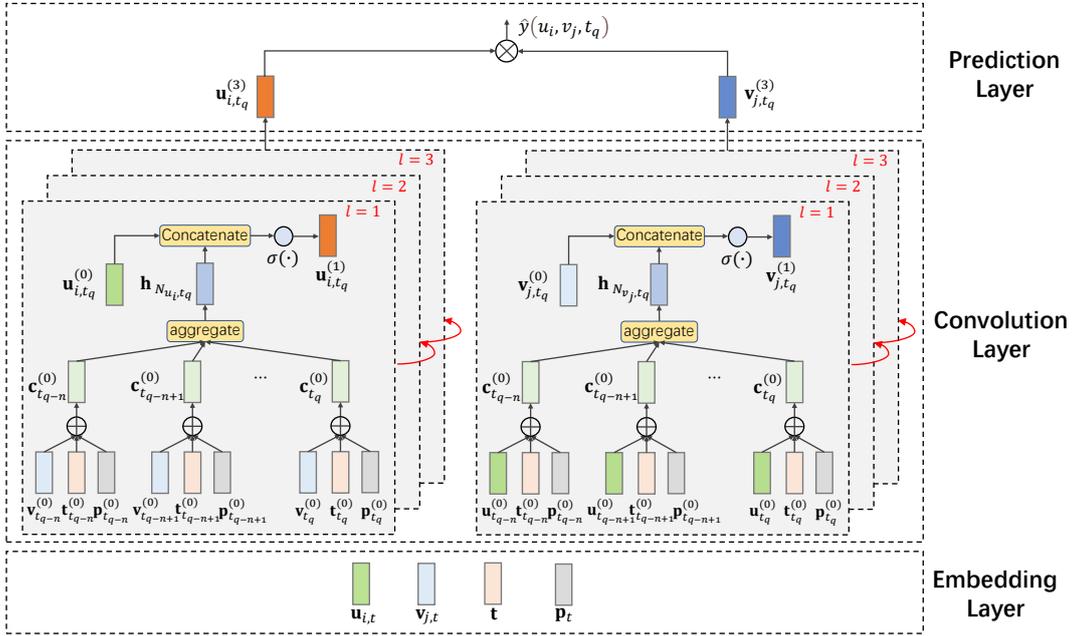

Fig. 2. The overall framework of PTGCN.

## 4.2 Embedding Layer

The embedding layer's goal is to map a given input to a low-dimensional vector representation. To model sequential patterns and temporal influence of interactions effectively, we encode user index, item index, interaction time, and the absolute position of each interaction in user neighborhood (or item neighborhood) into a shared latent space. In particular, we use user neighborhood to update the user embedding and item neighborhood to update the item embedding. Finally, we perform an add operation on them to obtain the hidden representation of each interaction.

**4.2.1 User embedding and item embedding**

First of all, we assume that the embedding of each user (or item) will change only when the interaction occurs. For all historical interactions, we create an embedding matrix $\mathbf{U} \in \mathbb{R}^{|U| \times d}$ for users and an embedding matrix $\mathbf{V} \in \mathbb{R}^{|V| \times d}$ for items, where $|U|$ and $|V|$ are the number of users and items, respectively, and $d$ is the latent dimension of embeddings. User embeddings and item embeddings are generated by model training and initialized randomly. For each interaction $i_{u_i, v_j, t}$, we perform a direct lookup operation on user and item indexes and obtain the user embedding $\mathbf{u}_{i,t}$ of user $u_i$ and the item embedding $\mathbf{v}_{j,t}$ of item $v_j$. Here, $\mathbf{u}_{i,t}$ and $\mathbf{v}_{j,t}$ denote the representation of $u_i$ and the representation of $v_j$, respectively, at time point $t$. Note that the user (or item) identifier is set to the value we want to look up.

### 4.2.2 Time embedding

Temporal information is essential to analyze individual interaction behaviors [9]. Because the sequential recommendation task is timing-dependent, we need to learn a proper time representation from the continuous-time nature of interactions. One of the most straightforward ways is to directly use the original feature values or transformations without embedding [49]. However, this method's performance is often poor because of the low capacity of representations. Besides, there are two methods to embed temporal information into low-dimensional vectors. The field embedding method learns a single field embedding for each numerical field by defining a continuous functional $\emptyset(\cdot)$ to map time intervals from the time domain to a $d$-dimensional vector space [11], [50]. The discretization method [51] converts numerical features into categorical ones using various heuristic discretization strategies and then assigns embeddings with the categorization strategy. In [51], the elapsed time is sliced into intervals whose length grows exponentially. For example, we can map the time in the range $[0, 1), [1, 2), [2, 4), \ldots, [2k, 2k + 1)$ into a categorical feature of $0, 1, 2, \ldots, k + 1$. Different groups of interactions may have different granularities of time slicing. Then, we perform a direct lookup on the categorical time features to obtain the time embedding $\mathbf{t}$ of time point $t$.

After comparing the effects of two commonly-used methods [50], [51] in the pre-experiment, we use the temporal encoding method proposed by Zhou *et al.* [51] in this study. For each interaction $i_{u_i,v_j,t}$ in $N_{u_i,t}$ or $N_{v_j,t}$, we can obtain the time embedding ($\mathbf{t}$) of time point $t$. Then, for each interaction $i_{u_i,v_j,t}$, we utilize the user neighborhood ($N_{u_i,t}$) of user $u_i$ at time point $t$ to obtain a new vector $\mathbf{u}_{i,t}$ and the item neighborhood ($N_{v_j,t}$) of item $v_j$ to obtain a new vector $\mathbf{v}_{j,t}$ at $t$. In particular, we leverage the time interval between two consecutive interactions in $N_{u_i,t}$ and time point $t$ to model the influence of historical interactions on the current state of $u_i$; also, this is similar to $v_j$.

### 4.2.3 Position embedding

To model sequential patterns effectively, we attempt to encode the position information (more specifically, the relative position) of a user-item interaction in the user neighborhood or item neighborhood according to the temporal information of each interaction. PTGCN employs the positional encoding approach proposed by Vaswani *et al.* [35], which does not introduce additional parameters and can extrapolate to sequence lengths longer than pre-defined fixed ones. The positional encodings have the same dimension as the input embeddings. As mentioned above, we select the latest $n$ neighbors before time point $t$ in a user-item interaction graph to constitute the user neighborhood and item neighborhood. In this setting, position embeddings contain specific information different from time embeddings. PTGCN then learns and updates the user and item embeddings at $t$ according to the user neighborhood and item neighborhood. In this way, for each interaction $i_{u_i,v_j,t}$, we can obtain the corresponding position embedding $\mathbf{p}_{i,t} \in \mathbb{R}^d$ of $u_i$ (or $\mathbf{p}_{j,t} \in \mathbb{R}^d$ of $v_j$).

## 4.3 Convolutional Layer

The convolutional layer builds upon the GCN architecture to model the sequential and temporal influence and capture high-order collaborative information in the bipartite graph structure. In addition, this layer refines the dynamic embeddings of users and items. To this end, we design a position-enhanced and time-aware graph convolution and the self-attention aggregator architecture and model the high-order connectivity by stacking multi-layer convolutions.

**4.3.1 Position-enhanced and time-aware graph convolution**

We integrate and update user and item embeddings using the graph convolution operation. The purpose of GCNs is to learn node representations by smoothing features over a graph [48]. A GCN iteratively performs the graph convolution operation to update each node's representation by aggregating its neighbors' features. Given the neighbor set of node $u$ ($N_u$) that contains its immediate neighbors at the $l$-th iteration, a GCN updates the node's embedding $\mathbf{h}_u^{(l)}$ by concatenating its current representation $\mathbf{h}_u^{(l-1)}$ and the aggregation of its neighboring feature vectors $\mathbf{h}_{N_u}^{(l)}$. Such a standard graph convolution can be formalized as

$$\mathbf{h}_{N_u}^{(l)} = \text{AGGREGATE}(\{\mathbf{h}_{u'}^{(l-1)} | u' \in N_u\}), \tag{1}$$

$$\mathbf{h}_u^{(l)} = \sigma(\mathbf{W}^{(l)} \cdot \text{CONCAT}(\mathbf{h}_u^{(l-1)}, \mathbf{h}_{N_u}^{(l)})), \tag{2}$$

where $\text{AGGREGATE}(\cdot)$ is an aggregation function, $\text{CONCAT}(\cdot)$ is the concatenation operation, $\mathbf{W}^{(l)}$ is a weight matrix, and $\sigma$ is a nonlinear activation function.

One fundamental limitation of most existing GCN models is that they failed to capture each node's position information in the neighborhood. In the scenario of sequential recommendations, we need to model the time effects of different interactions to obtain the dynamic representations of users and items. Therefore, we propose a position-enhanced and time-aware graph convolution by incorporating sequential and temporal information from interactions.

Given the neighborhood of $u_i$ at time point $t_q$ ($N_{u_i,t_q} = \{i_{u_i,v_j,t_m} | v_j \in V, q - n < m \leq q\}$), containing the latest $n$ interactions of $u_i$ before $t_q$, we update the user's embedding by aggregating the $n$ interactions in $N_{u_i,t_q}$. Assume that we have learned the parameters of the aggregator function (i.e., $\text{AGGREGATE}(\cdot)$), which aggregates the information from $N_{u_i,t_q}$. For all user-item interactions $\{i_{u_i,v_j,t} \in N_{u_i,t_q}\}$, we define the position-enhanced and time-aware graph convolution as

$$\mathbf{h}_{N_{u_i,t_q}}^{(l)} = \text{AGGREGATE}(\{(\mathbf{v}_{j,t}^{(l-1)}, \mathbf{t}, \mathbf{p}_{j,t}) | i_{u_i,v_j,t} \in N_{u_i,t_q}\}), \tag{3}$$

$$\mathbf{u}_{i,t_q}^{(l)} = \mathbf{W}_{U_2} \cdot \sigma(\mathbf{W}_{U_1} \cdot \text{CONCAT}(\mathbf{u}_{i,t_q}^{(l-1)}, \mathbf{h}_{N_{u_i,t_q}}^{(l)})), \tag{4}$$

where $\{(\mathbf{v}_{j,t}^{(l-1)}, \mathbf{t}, \mathbf{p}_{j,t})\}$ denotes the representations of the item nodes in $N_{u_i,t_q}$ at the $l$-th iteration. Due to the heterogeneity of nodes, different types of nodes have different feature spaces. For user nodes and item nodes in the bipartite graph, we design type-specific trainable transformation matrices $\mathbf{W}_{U_1}, \mathbf{W}_{V_1} \in \mathbb{R}^{d \times d}$ and $\mathbf{W}_{U_2}, \mathbf{W}_{V_2} \in \mathbb{R}^{2d \times d}$ to project the features of different types of nodes into the same feature space. The two types of matrices are shared by all graph convolutional layers.

Similarly, we can obtain the dynamic representation of each item in the same way.

$$\mathbf{h}_{N_{v_j,t_q}}^{(l)} = \text{AGGREGATE}(\{(\mathbf{u}_{i,t}^{(l-1)}, \mathbf{t}, \mathbf{p}_{i,t}) | i_{u_i,v_j,t} \in N_{v_j,t_q}\}, \tag{5}$$

$$\mathbf{v}_{j,t_q}^{(l)} = \sigma\left(\mathbf{W}_{V_2} \cdot \text{CONCAT}\left(\mathbf{W}_{V_1} \cdot \mathbf{v}_{j,t_q}^{(l-1)}, \mathbf{h}_{N_{v_j,t_q}}^{(l)}\right)\right). \tag{6}$$

Compared with existing GCNs, PTGCN redefines the neighborhood of nodes and integrates temporal and positional information into the graph convolution operation. As a result, the aggregation of the latest interactions can be implemented by various aggregator architectures, discussed in the following subsubsection.

**4.3.2 Aggregator architecture**

A few previous studies have specified the aggregator of GCNs, such as the weighted sum aggregator in [52] and mean aggregator in [48]. However, most GCN models neglect the order of nodes in a node's neighborhood, but this feature is essential to modeling sequential patterns for the sequential recommendation. As mentioned above, it is also necessary to simultaneously model the sequential influence and temporal influence. Considering that the self-attention mechanism has been applied for sequential recommendations and achieved remarkable results [5], [9], we design an aggregator using the self-attention mechanism. The key idea of the self-attention aggregator is to enrich each user feature and item feature with the corresponding time embedding and position embedding. The self-attention aggregator has $K$ identical nonlinear layers, each of which contains a self-attention layer, a feed-forward layer, and a vanilla attention layer. Next, we will introduce it with an example of the aggregation of user representations in detail.

***Self-attention layer***. For each $i_{u_i,v_j,t} \in N_{u_i,t_q}$, we can obtain an element $\mathbf{c}_j^{(l,k)}$ (or $\mathbf{c}_i^{(l,k)}$) in the $k$-th self-attention layer for each $v_j$ (or $u_i$) at the $l$-th iteration. Here, $\mathbf{c}_j^{(l,k)}$ is calculated as a weighted sum of item embedding, time embedding, and position embedding.

$$\mathbf{c}_j^{(l,k)} = \sum_{r=1}^n \alpha_{jr}^{(l,k)} \mathbf{c}_r^{(l,k-1)}, \tag{7}$$

$$\mathbf{c}_j^{(l,0)} = \mathbf{v}_{j,t}^{(l)} + \mathbf{t} + \mathbf{p}_{j,t}, \tag{8}$$

In Eq. (7), the weight coefficient $\alpha_{jr}^{(l,k)}$ is computed using a softmax function, defined as

$$\alpha_{jr}^{(l,k)} = \frac{\exp e_{jr}^{(l,k)}}{\sum_{r=1}^n \exp e_{jr}^{(l,k)}}, \tag{9}$$

where $e_{jr}^{(l,k)}$ is computed by an inner product operation that considers input, time, and position,

$$e_{jr}^{(l,k)} = \frac{\mathbf{c}_j^{(l,k-1)} \cdot (\mathbf{c}_r^{(l,k-1)})^T}{\sqrt{d}}, \tag{10}$$

and the scale factor $\sqrt{d}$ is used to penalize a higher inner product value, especially when the dimension is large.

***Point-wise feed-forward layer***. The self-attention layer models the sequential and temporal influences using a linear combination with adaptive weights. After each self-attention layer, we apply two linear transformations and a rectified linear unit (ReLU) activation between the two transformations, introducing non-linearity to the convolutional layer. In addition, we adopt layer normalization, residual connections, and dropout regularization techniques to avoid a few common problems caused by multi-layer neural networks, such as overfitting and vanishing gradients.

$$\mathbf{s}_j^{(l,k)} = FFN(\mathbf{c}_j^{(l,k)}) = \text{LayerNorm}(\text{Dropout}(\text{ReLU}(\mathbf{W}_1^{(k)} \cdot \mathbf{c}_j^{(l,k)})\mathbf{W}_2^{(k)}) + \mathbf{c}_j^{(l,k)}), \tag{11}$$

where $\text{ReLU}(\cdot)$ is the ReLU activation function, $\mathbf{W}_1^{(k)}, \mathbf{W}_2^{(k)} \in \mathbb{R}^{d \times d}$ denote two trainable parameter matrices shared by all graph convolutional layers, $\text{Dropout}(\cdot)$ is a dropout function, and $\text{LayerNorm}(\cdot)$ is the Layer normalization operation.

***Vanilla attention layer***. Given the user neighborhood at time point $t_q$, the output of interaction representations generated by the last feed-forward layer is a set $\{\mathbf{s}_j^{(l,K)}\}$, where $K$ is the number of nonlinear layers. A vanilla attention layer and the softmax function compute the normalized attention weight, similar to the self-attention layer.

$$\alpha_{ij}^{(l)} = \frac{\exp e_{ij}^{(l)}}{\sum_{j=1}^{n} \exp e_{ij}^{(l)}}, \tag{12}$$

$$e_{ij}^{(l)} = \frac{\mathbf{u}_{i,t_q}^{(l)} \cdot (\mathbf{s}_j^{(l,K)})^T}{\sqrt{d}}, \tag{13}$$

where $\alpha_{ij}^{(l)}$ denotes the attention weight between $\mathbf{u}_{i,t_q}^{(l)}$ and $\mathbf{s}_j^{(l,K)}$. After obtaining the attention weights in the vanilla attention layer, the aggregation representation of $N_{u_i,t_q}$ is calculated using the following equation:

$$\mathbf{h}_{N_{u_i,t_q}}^{(l)} = \sum_{i=1}^{n} \alpha_{ij}^{(l)} \mathbf{s}_j^{(l,K)}. \tag{14}$$

### 4.3.3 Stacking convolutions

The high-order connectivity used to model the collaborative signal is critical to evaluate users' item preferences. With the representations augmented by the first-order connectivity modeling, we can stack more position-enhanced and time-aware graph convolutions to capture the high-order connectivity in the bipartite graph. By stacking $l$ graph convolutions, a user (or an item) can receive the collaborative information from its $l$-hop neighbors.

---

**Algorithm 1**: Generating user embeddings

**Input**: Bipartite graph $G$, mini-batch set of users $\{u\}$, time point $t$, each user's set of recent interactions $N_{u,t}$, enumeration of node types $nodetypes = \{user: 0, item: 1\}$, depth $L$, type-specific transformation matrices $\mathbf{W}_{U_1}, \mathbf{W}_{V_1}$ and $\mathbf{W}_{U_2}, \mathbf{W}_{V_2}$, differentiable aggregator function $\text{AGGREGATE}(\cdot)$, neighborhood function $\mathcal{N}(interact, type)$ that returns the user neighborhood or item neighborhood according to the node type

**Output**: Vector representation $\mathbf{z}_{u,t}$ for each user $u$ at time $t$

01. $CS^{(L)} \leftarrow \cup_u N_{u,t}$ and $nodetype \leftarrow nodetypes[user]$;
02. **For** $l = L \dots 2$ **do**
03.     $CS^{(l-1)} \leftarrow CS^{(l)}$;
04.     **For** each $i_{u_i,v_j,t'}$ in $CS^{(l)}$ **do**
05.         $CS^{(l-1)} \leftarrow CS^{(l-1)} \cup \mathcal{N}(i_{u_i,v_j,t'}, nodetype)$;
06.     **End for**
07.     $nodetype \leftarrow 1 - nodetype$;
08. **End for**
09. **For** $l = 1 \dots L$ **do**
10.     **For** each $i_{u_i,v_j,t'}$ in $CS^{(l)}$ **do**
11.         **If** $nodetype == 0$ **then**
12.             Aggregate the representations of the item nodes in the user neighborhood using Eq. (3);
13.             Update the embedding of the target user $\mathbf{u}_{i,t'}^{(l)}$ using Eq. (4);
14.         **Else if** $nodetype == 1$ **then**
15.             Aggregate the representations of the user nodes in the item neighborhood using Eq. (5);
16.             Update the embedding of the target item $\mathbf{v}_{j,t'}^{(l)}$ using Eq. (6);
17.         **End if**
18.     **End for**
19.     $nodetype \leftarrow 1 - nodetype$;
20. **End if**
21. **Return** $\{\mathbf{z}_{u,t} \leftarrow \mathbf{u}_{i,t}^{(L)}\}$;

---

Here, we take user embedding as an example to illustrate how stacked convolutions generate user dynamic embeddings. *Algorithm 1* details the process of generating embeddings for a mini-batch set of users at a given time point. In the $l$-th iteration, we first compute the neighborhood of the target node according to the node type (user or item) and then apply the graph convolution to generate the $l$-th layer representation of the target node. For example, if the target node is a user node, the neighborhood function returns its item neighborhood in the bipartite graph. The output of the last convolutional layer is the final output embedding $\mathbf{z}_{u_i,t} = \mathbf{u}_{i,t}^{(L)}$. Similarly, we can use the same method to obtain dynamic item embeddings $\mathbf{z}_{v_j,t} = \mathbf{v}_{j,t}^{(L)}$.

### 4.4 Model Prediction

After performing stacked multi-layer convolutions, we obtain a user's representation at each historical interaction moment. We can utilize the representation for all historical interaction moments and the user's historical behavior sequence to predict future behavioral preferences. Therefore, for the target user $u_i$ and given time point $t_N$, we perform a forward multi-layer graph convolution operation to obtain the user's representation $\mathbf{z}_{u_i,t_N}$ and a target item's representation $\mathbf{z}_{v_j,t_N}$. Finally, a simple inner product operation is used to predict the user's preference for the target item:

$$\hat{y}(u_i, v_j, t_N) = (\mathbf{z}_{u_i,t_N})^T \cdot \mathbf{z}_{v_j,t_N}. \tag{15}$$

Because the PTGCN model is inductive, we use the user and item embeddings at historical moments to update the user and item embeddings at new interaction moments and then predict possible future interactions without retraining the model. However, we employ an inner product as the interaction function to implement user sequence learning in this work. Other more complex functions, such as neural-network-based interaction functions [53], are left to explore in our future work.

### 4.5 Optimization

---
**Algorithm 2**: Constructing training instances

**Input**: Bipartite graph $G$, enumeration of node types $nodetypes = \{user: 0, item: 1\}$, depth $L$, and neighborhood function $\mathcal{N}(interact, type)$

**Output**: a set of training instances $D$

01. $D \leftarrow \emptyset$ and $nodetype \leftarrow nodetypes[user]$;
02. **For** each $i_{u_i,v_j,t}$ in $G$ **do**
03.     Get a negative sample $v_j'$ for $v_j$ at time $t$ by the negative sampling method;
04.     $NF_{u_i,t} = CS_{u_i}^{(L)} \leftarrow N_{u_i,t}$, $NF_{v_j,t} = CS_{v_j}^{(L)} \leftarrow N_{v_j,t}$, and $NF_{v_j',t} = CS_{v_j'}^{(L)} \leftarrow N_{v_j',t}$;
05.     **For** $l = L \ldots 2$ **do**
06.         **For** each $i_{u',v',t'}$ in $CS_{u_i}^{(l)}$ **do**
07.             $CS_{u_i}^{(l-1)} \leftarrow CS_{u_i}^{(l)} \cup \mathcal{N}(i_{u',v',t'}, nodetype)$;
08.         **End for**
09.         Append $CS_{u_i}^{(l-1)}$ to $NF_{u_i,t}$;
10.         **For** each $i_{u',v',t'}$ in $CS_{v_j}^{(l)}$ **do**
11.             $CS_{v_j}^{(l-1)} \leftarrow CS_{v_j}^{(l)} \cup \mathcal{N}(i_{u',v',t'}, nodetype)$;
12.         **End for**
13.         Append $CS_{v_j}^{(l-1)}$ to $NF_{v_j,t}$;
14.         **For** each $i_{u',v',t'}$ in $CS_{v_j'}^{(l)}$ **do**
15.             $CS_{v_j'}^{(l-1)} \leftarrow CS_{v_j'}^{(l)} \cup \mathcal{N}(i_{u',v',t'}, nodetype)$;
16.         **End for**
17.         Append $CS_{v_j'}^{(l-1)}$ to $NF_{v_j',t}$;
18.         $nodetype \leftarrow 1 - nodetype$;
19.     **End for**
20.     Add a training instance $s_{u_i,v_j,t} = <NF_{u_i,t}, NF_{v_j,t}, NF_{v_j',t}>$ to $D$;
21. **End for**
22. **Return** $D$;

---

The goal of our model is to provide a list of top-$k$ ranked items for the target user. Considering that user interactions are always implicit, the negative sampling technique used in [9] is adopted to optimize the ranking of items. For a specific positive output $\hat{y}(u_i, v_j, t)$, we sample one negative item $v'_j$ at time point $t$. Thus, we can build a training instance $s_{u_i,v_j,t} = <NF_{u_i,t}, NF_{v_j,t}, NF_{v'_j,t}>$ for each interaction $i_{u_i,v_j,t}$, including the user node flow $NF_{u_i,t}$, the item node flow $NF_{v_j,t}$, and the item node flow $NF_{v'_j,t}$ of the negative (or called unobserved) item $v'_j$ at $t$. *Algorithm 2* presents the construction process of training instances.

We train PTGCN in an unsupervised manner using a binary cross-entropy as the loss function. Considering that our goal is to predict an item in which the target user is interested at a specific moment, we define the loss function terms over the interactions based on the dynamic embeddings of users and items. For all historical user interactions, the purpose of the training process is to optimize the PTGCN's parameters so that the output embeddings of $u_i$ and $v_j$ at time $t$ for each interaction are close together and the difference between the embeddings of $u_i$ and $v'_j$ is as large as possible. For the training set $D = \{s_{u_i,v_j,t}\}$, we can define the loss function as

$$-\sum_{s_{u,v,t} \in D} \left[ \log\left(\sigma\left(\hat{y}(u_i, v_j, t)\right)\right) + \log\left(1 - \sigma\left(\hat{y}(u_i, v'_j, t)\right)\right) \right] + \lambda \|W\|, \quad (16)$$

where $\sigma(\cdot)$ is the sigmoid function, $W$ is the set of embedding matrices, $\|\cdot\|$ denotes the Frobenius norm, and $\lambda$ is the regularization parameter.

Adam [54], the standard stochastic optimization method, is used to optimize the objective function in this work. Since each training sample $s_{u_i,v_j,t}$ can be constructed independently, we apply the mini-batch stochastic gradient descent (SGD) method to speed up the training process. For the details of the whole training process of PTGCN, please refer to *Algorithm 3*.

**Algorithm 3**: Training PTGCN
**Input**: training set $D$
**Output**: the parameter set of PTGCN $\Theta$

01. Initialize $\Theta$;
02. **While** (exceed(maximum number of iterations) == FALSE) **do**
03.     Randomly select a batch of training instances $D_b$ from $D$;
04.     **For** each $i_{u_i,v_j,t}$ in $D_b$ **do**
05.         Calculate the representations of $z_{u_i,t}$, $z_{v_j,t}$, $z_{v'_j,t}$;
06.     **End for**
07.     Find $\Theta$ minimizing the objective function (Eq. (16)) with $D_b$;
08. **End while**
09. **Return** $\Theta$

## 5 Experiment and Result Analysis

### 5.1 Research Questions

The experiments were conducted by answering the following four research questions:

**RQ1**: *Can PTGCN outperform state-of-the-art baselines for sequential recommendation tasks?*

**RQ2**: *Is modeling temporal dynamics of items beneficial to the sequential recommendation?*

**RQ3**: *Does the high-order connectivity contribute to better recommendation performance?*

### 5.2 Datasets

We compared our method with selected baselines on three widely used datasets from two real-world platforms (i.e., MovieLens and Amazon). These datasets are publicly available on the Internet and have different domains, sizes, and sparsity.

- MovieLens is a stable benchmark dataset widely used for recommendation algorithm

evaluation. The version of MovieLens-1M[1] that includes 1 million movie ratings is used to evaluate sequential recommendation algorithms in our experiment.

- Amazon [2] is a large-scale dataset obtained from Amazon review datasets [55], [56], comprising large corpora of product reviews crawled from the Amazon website. We selected two categories, CDs_and_Vinyl and Movies_and_TV, from the original dataset.

The three datasets contain user-item interactions, each of which records user ID, item ID, rating (or review), and the corresponding interaction timestamp. Reviews or ratings are treated as implicit feedback, i.e., the interactions between users and items. All the interactions are sorted in the chronological order of timestamps. For the MovieLens dataset, we used the 5-core setting [8], [9] to filter out cold-start users and items with fewer than five interactions. We adopted a different 10-core setting to ensure that each user or item has at least ten interactions for the Amazon datasets, which are much sparser. As with [5], [9], each user's last interaction was selected for testing, and all previous interactions were used to build the training set. Table 2 shows the statistics of the experimental datasets.

Table 2. Statistics of the experimental datasets.

| Dataset | #Users | #Items | #Interactions | Avg. Interactions per user | Avg. Interactions per item |
|---|---|---|---|---|---|
| MovieLens | 6,040 | 3,416 | 999,611 | 165.5 | 292.6 |
| Amazon CDs_and_Vinyl | 17,965 | 14,253 | 444,285 | 24.7 | 31.2 |
| Amazon Movies_and_TV | 84,115 | 30,881 | 1,890,004 | 22.5 | 61.2 |

## 5.3 Experimental Setups
### 5.3.1 Evaluation Metrics

To evaluate the recommendation performance of PTGCN and baselines, we employ two commonly used metrics: $Recall@k$ ($R@k$) and normalized discounted cumulative gain@k ($NDCG@k$), where $k$ is set to five or ten. In the sequential recommendation task, $R@k$ has a strong positive correlation with $Precision@k$ ($P@k$) and $F1@k$ because there is only one positive item at a time of interaction. Compared with $P@k$ and $F1@k$, $R@k$ can show the recommendation ability to find out more candidate items. Therefore, we do not consider $P@k$ and $F1@k$ in this study.

$R@k$ evaluates how many candidate items in the top-$k$ recommendation hit the actual items in the test set.

$$R@k = \frac{1}{N}\sum_{u=1}^{N} R_u@k = \frac{1}{N}\sum_{u=1}^{N} \frac{|S_u(k) \cap V_u|}{|V_u|}, \qquad (17)$$

where $S_u(k)$ is a set of candidate items in the top-$k$ recommendation to user $u$ and $V_u$ denotes a set of actual items with which the user interacts at a given time in the test set. Note that $|V_u| = 1$.

$NDCG@k$ measures the ranking performance of a recommendation algorithm by considering the order of actually relevant items.

$$NDCG@k = \frac{1}{N}\sum_{u=1}^{N} \frac{1}{Z_u}\sum_{j=1}^{k} \frac{2^{I(|\{s_u^j\} \cap V_u|)}-1}{log_2(j+1)}, \qquad (18)$$

where $s_u^j$ is the $j$-th recommended item in $S_u(k)$, $I(\cdot)$ denotes the indicator function, and $Z$ is a normalized constant, i.e., the maximum value of $DCG@k$.

### 5.3.2 Baselines

To demonstrate the effectiveness of PTGCN, we compared it with a few competitive baseline approaches providing source code available on the Internet. These baselines include one classic recommendation method without considering sequential patterns and eight sequential methods using different recommendation techniques, listed as follows.

---
[1] https://grouplens.org/datasets/movielens/1m/
[2] https://nijianmo.github.io/amazon/index.html

*Non-sequential method*:
- BPR (Rendle *et al.*, 2009) [57]. This method is a classic method based on Bayesian personalized ranking (BPR) for the top-$k$ item recommendation. It models the order of candidate items by a pairwise ranking loss without considering sequential patterns.

*Sequential method - first-order Markov chain-based approach*:
- FPMC (Rendle *et al.*, 2010) [1]. This method is a classic method based on factorized personalized Markov chains for next-basket recommendations. It utilizes matrix factorization and first-order Markov Chains to capture long-term user preferences and dynamic transitions of sequential behavior, respectively.

*Sequential method - RNN-based approach*:
- GRU4Rec+ (Hidasi and Karatzoglou, 2018) [6]. This method is an improved version of GRU4Rec [58], an RNN-based approach for session-based recommendations. In particular, GRU4Rec+ adopts a different ranking loss function and sampling strategy. As a result, it shows significant improvement over GRU4Rec.

*Sequential method - CNN-based approach*:
- Caser (Tang and Wang, 2018) [8]. This method is a sequential recommendation method based on convolutional sequence embedding. It embeds a sequence of recent items into an "image" and models sequential influence as local features of the "image" with convolutional filters.

*Sequential method - Self-attentive approaches*:
- SASRec (Kang and McAuley, 2018) [5]. This method is proposed based on Transformer. It only utilizes a self-attention mechanism without the RNN architecture to capture sequential patterns and an attention mechanism to predict based on relatively few actions.
- BERT4Rec (Sun *et al.*, 2019) [36]. This method is a sequential recommendation method based on BERT [59]. Compared with SASRec and RNN-based methods using the left-to-right unidirectional architecture, BERT4Rec leverages bidirectional self-attention to model users' behavior sequences.
- TiSASRec (Li *et al.*, 2020) [9]. This method is an improved version of SASRec. Compared with SASRec, TiSASRec uses both the absolute positions of interactions and the time intervals between different interactions to improve sequential recommendation performance.

*Sequential method - GCN-based approaches*:
- DGCF (Li *et al.*, 2020) [60]. This method is a bipartite graph-based dynamic recommender system that captures collaborative and sequential relations of items and users and updates user and item embeddings simultaneously using three update mechanisms. It fuses the high-order connectivity of users and items with a simple multi-layer perceptron (MLP) and does not consider time and order information.
- DHCN (Xia *et al.*, 2021) [43]. This method is a session-based recommendation method based on the GCN architecture. It introduces self-supervised learning to train a hypergraph convolutional network to improve next-item recommendation performance.

Although the proposed and baseline approaches aim to predict the probability of an item being recommended to the target user, they differ in representing users and items. These approaches are assumed to have the same dimension of hidden variables (denoted by $d$), the same length of sequences (denoted by $s$), and the same size of samples (denoted by $n$). We approximate each approach's time complexity by calculating user and item representations for simplicity. Table 3 shows the time complexity of PTGCN and the baselines. PTGCN has the same order of magnitude as SASRec, BERT4Rec, and TiSASRec in time complexity because $s$ is usually smaller than $d$, and its time complexity is lower than those of Caser, DHCN, and DGCF.

For each baseline approach, the hyper-parameters were configured according to the default settings declared in the corresponding paper (see Subsection 5.3.3), and the trainable parameters were fine-tuned using the validation set. If the performance on the validation set does not improve after 50 epochs, the training process will be terminated.

**5.3.3 Experiment Settings**

All experiments were carried out on a Lenovo ThinkStation P910 workstation with dual processors (Intel Core i9-7920X, 2.9 GHz) and one graphics processing unit (NVIDIA GeForce GTX 1080Ti, 12GB). The operating system of the workstation was Microsoft Windows 10 (64-bit). Our experiment's implementation programs were coded with Python 3.8, and the deep learning framework we employed was PyTorch[3] 1.8.

The embedding dimension and batch size were set to 160 and 64, respectively, for PTGCN and all the baselines. The learning and dropout rates were set to 1e-4 and 0.1, respectively, when training PTGCN on the experimental datasets. Sample instances were randomly split into the training set, validation set, and test set for the three datasets. More specifically, 80% of sample instances were used for training, 10% for validation, and the remaining 10% for testing. As with most graph convolution methods [61], the depth of PTGCN was set to two to model the high-order collaborative signal. The maximum sequence lengths for the two convolutional layers were set to 50 and 20, respectively. For each aggregator, the number of self-attention layers was set to one; the number of identical nonlinear layers was set to six; the number of attention heads in the self-attention layer was set to eight. For more details of our method, please refer to the implementation source code available at https://github.com/drhuangliwei/PTGCN.

The seven deep learning-based baselines' settings are introduced below. We report the best result of each baseline under its optimal setting on the validation set.

- GRU4Rec+[4]: The number of hidden layers in the GRU4Rec+ model was set to 100. In the model training process, we optimized the BPR loss function with a learning rate of 0.05 and a momentum of 0.2, and the dropout rate was set to 0.2.
- Caser[5]: The Markov order $L$ and the target number $T$ were set to nine and three, respectively. For each height $h$, the number of horizontal filters was 16, and the number of vertical filters was four. Besides, the number of latent dimensions, the sequence length, and the number of negative samples were set to 50, 5, and 3, respectively. The Caser model was trained using Adam with a learning rate of 1e-4 and a dropout rate of 0.5 for fully-connected layers.
- SASRec[6]: The SASRec model with two self-attention blocks was trained using Adam with a learning rate of 1e-3. The dropout rate and the maximum sequence length were set to 0.2 and 200, respectively.
- BERT4Rec[7]: The BERT4Rec model was trained using Adam with a learning rate of 1e-4, $\beta_1$ = 0.9, $\beta_2$ = 0.999, $\ell_2$ regularization of 0.01, and linear decay of the learning rate. The gradient was clipped after its $\ell_2$ norm reached a threshold of five. The layer number, the head number, the maximum sequence length, the dimensionality of each head, and the mask proportion were set to 2, 2, 200, 32, and 0.2, respectively.

---
[3] https://pytorch.org/
[4] https://github.com/hidasib/GRU4Rec
[5] https://github.com/graytowne/caser
[6] https://github.com/kang205/SASRec
[7] https://github.com/FeiSun/BERT4Rec

- TiSASRec[8]: The TiSASRec model with two self-attention layers was trained using Adam with a learning rate of 1e-3. The dropout rate and the maximum sequence length were set to 0.2 and 50, respectively. Besides, the max time intervals on the MovieLens and Amazon datasets were 2,048 and 512, respectively, and the $\ell_2$ regularizations for the three datasets were selected from {0, 5e-5}.
- DGCF[9]: The DGCF model was trained using Adam with a learning rate of 1e-3 and an $\ell_2$ penalty of 1e-3. The smoothing coefficients $\lambda$ and $\alpha$ in loss function were set to one.
- DHCN[10]: The DHCN model was trained using Adam with an initial learning rate of 1e-3 and an $\ell_2$ regularization of 1e-5. The number of layers was selected from {1, 2, 3} for the three different datasets.

## 5.4 Results
**5.4.1 Performance Comparison (RQ1)**
*5.4.1.1 Overall result*

To answer **RQ1**, Table 4 shows the overall performance of the ten approaches on the three datasets. The underlined numbers stand for the best result of the baselines, and the numbers shown in bold represent the best result in each row. BPR utilizes only matrix factorization to model user preference, which cannot capture sequential patterns in user behavior. As a result, it performed the worst among the seven baselines. Although FPMC, a Markov chain-based method, can model the sequential influence, they cannot model the long-term effect of user behavior well. Therefore, it obtained sub-optimal results compared with the neural network-based approach. Among all the neural network-based baselines (i.e., GRU4Rec+, Caser, SASRec, BERT4Rec, DHCN, TiSASRec, and DGCF), DGCF, a SOTA approach, achieved the best performance because it is a dynamic recommendation method that can update the representations of users and items at any time and utilize the first- and second-order interactions to obtain the representations of users and items.

PTGCN outperformed the best baseline, DGCF, on the three datasets in terms of the two metrics. Moreover, the last column of Table 4 indicates a substantial performance improvement of PTGCN over DGCF. The main reasons may include the following three aspects. First, although DGCF also utilizes the second-order interactions to update the representations of users and items, it only uses the most recent interaction to obtain the dynamic representations of users and items in each step of the updating process. Second, DGCF does not update the node representations of the first-order neighborhood with the second-order interaction information in the same way as GNNs. It is, therefore, hard to mine the relationship between nodes in the first- and second-order neighborhoods. Third, DGCF does not take advantage of the time and position information to model the sequential patterns of user behavior. Hence, PTGCN achieved the best recommendation results on the three datasets.

*5.4.1.2 Recommendation for cold-start users*

The data sparsity problem is one of the critical issues that affect the performance of recommender systems. It is challenging to capture the dynamic preferences of users with few interactions (or called cold-start users). This study attempts to alleviate this problem by exploiting the high-order connectivity and modeling temporal dynamics of items. To test the recommendation performance of PTGCN for cold-start users, we grouped such users in the test set according to the number of historical interactions of each user. More specifically, the number of user groups was set to four for the three datasets. In the MovieLens dataset, cold-start users were divided into four groups with interactions less than 20, 30, 40, and 50. Since the average number of interactions per user in the other two Amazon datasets is about 20, we set four groups of cold-start users whose interactions were less than 15, 20, 25, and 30. Note that $k$ was set to ten in this experiment.

---

[8] https://github.com/JiachengLi1995/TiSASRec
[9] https://github.com/CRIPAC-DIG/DGCF
[10] https://github.com/xiaxin1998/DHCN

Fig. 3 presents the recommendation performance of the eight neural network-based methods for different user groups. The X-axis denotes different user groups, and the left and right Y-axes represent the number of users in a group and an evaluation metric, respectively. As shown in Fig. 3, PTGCN achieves consistent advantages over the other seven baseline methods for all the user groups with different sparsity levels. It is worth noting that the performance improvements in the first two groups are more significant than those of the other ones in terms of $NDCG@10$ and $R@10$. For example, compared with TiSASRec that does not consider the high-order collaborative information and temporal dynamics of items, the $NDCG@10$ values of PTGCN were increased by 40.79% and 28.54% for the first and second groups, respectively, on the MovieLens dataset. This finding is also valid for the other two Amazon datasets. Such a result indicates that modeling the high-order connectivity and temporal dynamics of items is beneficial to improving sequential recommendations for cold-start users. Hence, it is very promising for our method to solve the user-item interaction sparsity problem in the sequential recommendation task.

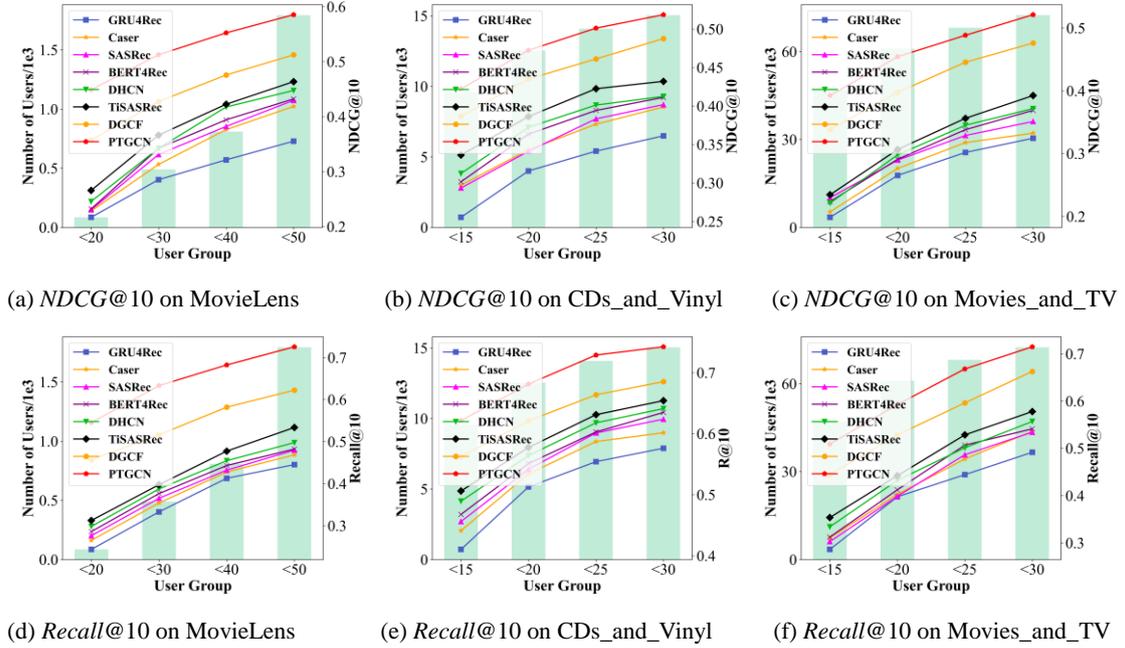

(a) $NDCG@10$ on MovieLens  (b) $NDCG@10$ on CDs_and_Vinyl  (c) $NDCG@10$ on Movies_and_TV

(d) $Recall@10$ on MovieLens  (e) $Recall@10$ on CDs_and_Vinyl  (f) $Recall@10$ on Movies_and_TV

Fig. 3. Performance comparison over the sparsity distribution of user groups.

Table 3. Comparison among different methods in time complexity.

| | FPMC | GRU4Rec+ | Caser | SASRec | BERT4Rec | DHCN | TiSASRec | DGCF | PTGCN |
|---|---|---|---|---|---|---|---|---|---|
| Markov chain | ✓ | | | | | | | | |
| RNN | | ✓ | | | | | | | |
| CNN | | | ✓ | | | | | | |
| Self-attention | | | | ✓ | ✓ | | ✓ | | |
| GCN | | | | | | ✓ | | ✓ | ✓ |
| Time | $O(nd)$ | $O(nd^2)$ | $O(nshd^2)$ | $O(nks^2d)$ | $O(nks^2d)$ | $O(nks^2d)$ | $O(nks^2d)$ | $O(nsd^2)$ | $O(ns^ld)$ |

Note: $h$ is the height of a horizontal filter in Caser, $k$ is the number of self-attention layers, and $l$ is the level of collaborative signals.

Table 4. Comparison among different methods in recommendation performance.

| Dataset | Metrics | BPR | FPMC | GRU4Rec+ | Caser | SASRec | BERT4Rec | DHCN | TiSASRec | DGCF | PTGCN | Improvement |
|---|---|---|---|---|---|---|---|---|---|---|---|---|
| MovieLens | $R@5$ | 0.4012 | 0.4431 | 0.5075 | 0.5758 | 0.5834 | 0.5946 | 0.6012 | 0.6246 | <u>0.6827</u> | **0.7651** | 12.07% |
| | $NDGG@5$ | 0.2901 | 0.3332 | 0.3913 | 0.4436 | 0.4562 | 0.4658 | 0.4731 | 0.4887 | <u>0.5269</u> | **0.5994** | 13.86% |
| | $R@10$ | 0.5432 | 0.5763 | 0.6345 | 0.7064 | 0.7124 | 0.7215 | 0.7348 | 0.7591 | <u>0.7912</u> | **0.8746** | 10.54% |
| | $NDGG@10$ | 0.3342 | 0.3741 | 0.4252 | 0.4813 | 0.4871 | 0.4974 | 0.5012 | 0.5235 | <u>0.5673</u> | **0.6351** | 11.95% |
| Amazon CDs_and_Vinyl | $R@5$ | 0.3674 | 0.4034 | 0.4661 | 0.5128 | 0.5125 | 0.5235 | 0.5346 | 0.5537 | <u>0.5824</u> | **0.6421** | 10.25% |
| | $NDGG@5$ | 0.2683 | 0.2854 | 0.3245 | 0.3762 | 0.3716 | 0.3912 | 0.3812 | 0.4034 | <u>0.4389</u> | **0.4946** | 12.69% |
| | $R@10$ | 0.5218 | 0.5514 | 0.6012 | 0.6348 | 0.6438 | 0.6542 | 0.6645 | 0.6824 | <u>0.6917</u> | **0.7686** | 11.12% |
| | $NDGG@10$ | 0.3248 | 0.3552 | 0.3762 | 0.4189 | 0.4127 | 0.4238 | 0.4213 | 0.4486 | <u>0.4923</u> | **0.5342** | 8.51% |
| Amazon Movies_and_TV | $R@5$ | 0.3654 | 0.3879 | 0.4426 | 0.4782 | 0.4882 | 0.4986 | 0.5083 | 0.5293 | <u>0.6127</u> | **0.6784** | 10.72% |
| | $NDGG@5$ | 0.2458 | 0.2762 | 0.3012 | 0.3389 | 0.3528 | 0.3636 | 0.3619 | 0.3881 | <u>0.4764</u> | **0.5341** | 12.11% |
| | $R@10$ | 0.5042 | 0.5423 | 0.5829 | 0.6281 | 0.6139 | 0.6239 | 0.6349 | 0.6482 | <u>0.7347</u> | **0.8012** | 9.05% |
| | $NDGG@10$ | 0.2979 | 0.3349 | 0.3689 | 0.3992 | 0.4071 | 0.4213 | 0.4118 | 0.4267 | <u>0.5236</u> | **0.5722** | 9.28% |

### 5.4.2 Impact of Temporal Dynamics of Items (RQ2)

**RQ2**'s goal is to analyze the impact of modeling the temporal dynamics of items on the accuracy of sequential recommendations. In this experiment, the item embedding was designated as a static value to learn from input data. Considering that an item's state does not change over time, we model only the first-order collaborative information of users. Such a model, similar to TiSASRec, is denoted as PTGCN-1-user. To compare with PTGCN-1-user, we designed a simplified version of PTGCN, PTGCN-1 (depth $L = 1$), which simultaneously models the temporal dynamics of users and items by utilizing only the first-order connectivity between users and items.

As shown in Table 5, PTGCN-1 performs better than PTGCN-1-user in terms of the two evaluation metrics. Specifically, compared with PTGCN-1-user, the $R@5$, $NDGG@5$, $R@10$, and $NDGG@10$ values of PTGCN-1 on the three datasets were increased, on average, by 8.99%, 11.50%, 6.52%, and 8.92%, respectively. This result further demonstrates the superiority of introducing the temporal dynamics of items to the sequential recommendation, which is one of the leading technical contributions of this study, over existing approaches that only consider user dynamics (or preference).

Table 5. Impact of temporal dynamics of items on recommendation performance.

| Dataset | Model | $R@5$ | $NDGG@5$ | $R@10$ | $NDGG@10$ |
| --- | --- | --- | --- | --- | --- |
| MovieLens | PTGCN-1 | 0.6873 | 0.5128 | 0.7984 | 0.6432 |
|  | PTGCN-1-user | 0.6345 | 0.4674 | 0.7643 | 0.5876 |
| Amazon CDs_and_Vinyl | PTGCN-1 | 0.5876 | 0.4336 | 0.7125 | 0.4764 |
|  | PTGCN-1-user | 0.5332 | 0.3876 | 0.6774 | 0.4432 |
| Amazon Movies_and_TV | PTGCN-1 | 0.6017 | 0.4659 | 0.7342 | 0.4976 |
|  | PTGCN-1-user | 0.5547 | 0.4126 | 0.6679 | 0.4532 |

### 5.4.3 Impact of High-order Connectivity (RQ3)

**RQ3**'s goal is to analyze the impact of high-order connectivity on the performance of sequential recommendations. To this end, we tested the performance of PTGCN with three different levels of collaborative signals (or different values of depth $L$), namely PTGCN-1, PTGCN-2, and PTGCN-3. Table 6 presents the comparison results on the three datasets. The numbers shown in bold represent the best result. PTGCN-0 denotes a PTGCN model without graph convolutions, similar to a matrix factorization-based model. According to Table 6, we have the following observations:

- Generally speaking, increasing the depth of PTGCN can improve the recommendation performance. Because PTGCN-0 does not consider sequential patterns of users and items, it degenerates into a neural network-based collaborative filtering model [53]. Therefore, its recommendation performance was the worst among the four models. Compared with PTGCN-1, which only considers the first-order neighborhood in the bipartite graph, PTGCN-2 achieved consistent performance improvements over PTGCN-1 across all the datasets.
- When stacking more graph convolutional layers on the top of PTGCN-2, we find that the overfitting problem for PTGCN-3 occurred on the three datasets. Although PTGCN-3 performed slightly worse than PTGCN-2, it still outperformed PTGCN-1. This problem happened because a deep architecture of graph convolutional layers may bring noise data to the representation learning of user features and item features. This result is also consistent with previous studies that stacking multiple GCN layers will result in over-smoothing [61]; that is to say, all vertices on the bipartite graph will converge to the same value. Therefore, setting two GCN layers for PTGCN is sufficient to capture the high-order connectivity.

Table 6. Impact of high-order connectivity on recommendation performance.

| Dataset | Model | $R@5$ | $NDGG@5$ | $R@10$ | $NDGG@10$ |
|---|---|---|---|---|---|
| MovieLens | PTGCN-0 | 0.4634 | 0.4014 | 0.6548 | 0.4235 |
|  | PTGCN-1 | 0.5557 | 0.4301 | 0.7012 | 0.4573 |
|  | PTGCN-2 | **0.7651** | **0.5994** | **0.8746** | **0.6351** |
|  | PTGCN-3 | 0.7458 | 0.5834 | 0.8623 | 0.6237 |
| Amazon CDs_and_Vinyl | PTGCN-0 | 0.4437 | 0.3326 | 0.5537 | 0.3546 |
|  | PTGCN-1 | 0.5126 | 0.3756 | 0.6342 | 0.4326 |
|  | PTGCN-2 | **0.6421** | **0.4946** | **0.7686** | **0.5342** |
|  | PTGCN-3 | 0.6348 | 0.4865 | 0.7534 | 0.5215 |
| Amazon Movies_and_TV | PTGCN-0 | 0.4537 | 0.3352 | 0.5876 | 0.3657 |
|  | PTGCN-1 | 0.5432 | 0.3875 | 0.6457 | 0.4368 |
|  | PTGCN-2 | **0.6784** | **0.5341** | **0.8012** | **0.5722** |
|  | PTGCN-3 | 0.6654 | 0.5213 | 0.7895 | 0.5576 |

## 5.5 Discussion

### 5.5.1 Impact of Data Partitioning

In machine learning, the amount of training data may affect a model's representation ability, and a small-size dataset may cause an overfitting problem. We first analyzed the impact of data partitioning on model performance to test the expression ability of the PTGCN model on training sets of different sizes. More specifically, we defined four commonly-used split ratios for the training, validation, and test sets, i.e., 50/25/25, 60/20/20, 70/15/15, and 80/10/10, and carried out the corresponding experiments on the three datasets. Fig. 4 shows the change in model performance with different data split ratios. The X-axis denotes four different split ratios, and the Y-axis represents the value of an evaluation metric. We can find from Fig. 4 that all the four metrics for model performance increase, with a growing proportion of 10% for the training set, suggesting that increasing the amount of training data can improve PTGCN's recommendation performance.

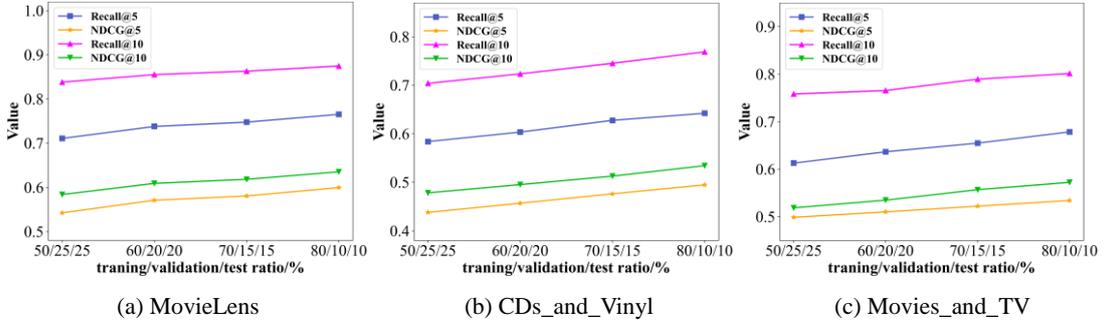

(a) MovieLens  (b) CDs_and_Vinyl  (c) Movies_and_TV

Fig. 4. Model performance change with different training/validation/test split ratios.

### 5.5.2 Sensitive Analysis of Parameters

*5.5.2.1 Embedding size*

The embedding dimension impacts the recommendation performance of PTGCN. Generally speaking, the higher the dimension of embeddings processed by a model, the stronger the model's representation ability [62]. However, at the same time, excessively high embedding dimensions often cause the overfitting problem. Fig. 5 presents the impact of this parameter on the recommendation performance of PTGCN on the three datasets. When the dimension of embeddings processed by PTGCN is smaller than 160, the $R@10$ and $NDCG@10$ values increase with the dimension of embeddings, indicating that the expressive ability of PTGCN is improved gradually. However, when the embedding dimension exceeds 160, the $R@10$ and $NDCG@10$ values start to decrease, implying that the model may encounter overfitting. Therefore, the number of embedding dimensions of PTGCN was set to 160 in our experiments.

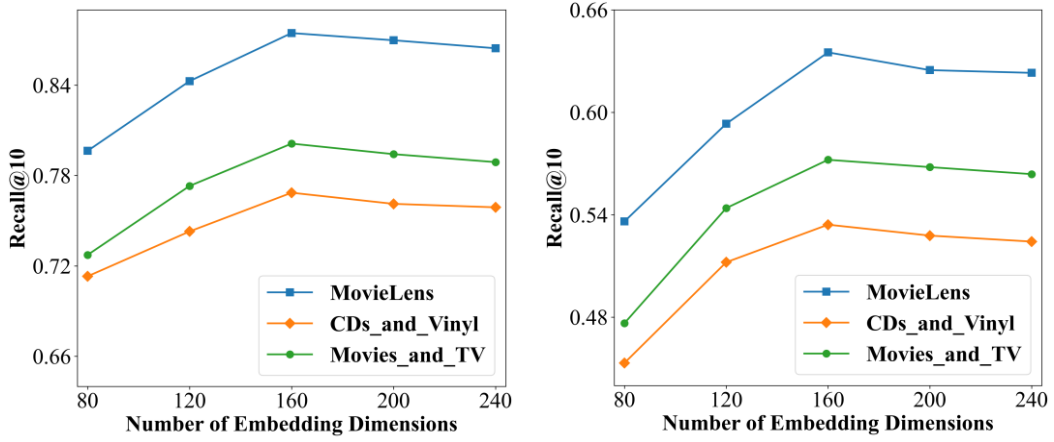

Fig. 5. Performance tuning with different embedding dimensions.

*5.5.2.2 Number of self-attention layers*

Generally speaking, if the PTGCN aggregator has more self-attention layers, the model will have a powerful expressive ability and thus obtain better recommendation performance. However, at the same time, stacking more self-attention layers will increase the computation load of the model. We need to make a trade-off between expressive ability and computation load in our experiment. Due to the limitation of hardware resources, the maximum number of self-attention layers was set to eight. Fig. 6 shows the impact of this parameter on the recommendation performance of PTGCN on the three datasets.

As shown in Fig. 6, the model obtains the best recommendation result when the aggregator has only one self-attention layer. As the number of self-attention layers increases, the model's performance remains nearly unchanged. In particular, with the increase of self-attention layers, there is a decline in recommendation performance on the MovieLens dataset. The reasons for this unexpected result may include two aspects. First, stacking multiple graph convolutional layers is, in essence, equivalent to deeply modeling the correlation between interactions. If each convolutional layer in PTGCN contains many self-attention layers, the model will become more complex, possibly leading to overfitting. Second, experimental data is insufficient for training complex models with more self-attention layers. The overfitting problem may also occur in this case.

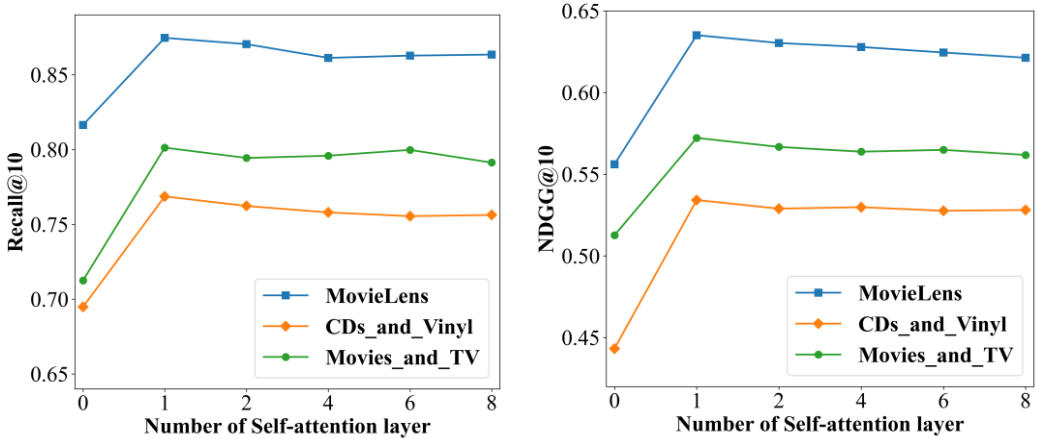

Fig. 6. Performance tuning with different self-attention layers.

### 5.5.3 Model Training Efficiency

*5.5.3.1 Training time per batch*

We also analyzed the training efficiency of eight methods based on neural networks, i.e., GRU4Rec+, Caser, SASRec, BERT4Rec, DHCN, TiSASRec, DGCF, and PTGCN. For the eight methods, the batch size and the embedding dimension were set to 64 and 160, respectively, to make a fair comparison under the same setting. The sequence lengths of the baseline methods were set according to the default values in the corresponding papers. In addition to the number of model parameters (denoted as #Params), we calculated the time of training one batch under the same experimental environment for each of the three datasets. Table 7 presents a detailed comparison among the eight methods regarding training time per batch on the three datasets. The numbers shown in bold represent the best result in each row, and the underlined numbers stand for the second-best result in the corresponding row.

DGCF has the maximum number of model parameters, while DHCN has the minimum number of model parameters. The number of PTGCN's parameters is 2.5 times that of DHCN; in other words, PTGCN is more complex than DHCN. Therefore, it needs more GPU memory to store deep learning models than SASRec, BERT4Rec, DHCN, and TiSASRec in the training process. It is worth noting that one standard GPU card with 3GB memory is sufficient to train the PTGCN models. Even then, our method has the third-fastest training speed regarding training time per batch. Because GRU4Rec+ utilizes GRU to update the embeddings of items, the sequence length in each batch was set to one during the model training process. As a result, it achieved the best training efficiency. However, its recommendation performance was the worst among the eight methods. DGCF can obtain suboptimal recommendation performance and training efficiency on the three datasets. Compared with DGCF, our method achieved better model performance while spending just a bit more training time per batch (the reasons refer to Subsubsubsection 5.4.1.1). Therefore, this experimental result indicates that PTGCN can better balance recommendation performance and training efficiency than the other seven methods.

*5.5.3.2 Convergence analysis*

As mentioned above, PTGCN uses Adam [54] to optimize the objective function. After calculating the training time per batch, we further analyzed the convergence of PTGCN when it was trained on the three datasets. The maximum epoch of model training and the learning rate were set to 50 and 1e-4, respectively. The patience parameter specifies the number of iterations where the loss on the training set is greater than or equal to the previously lowest loss before the model training process ends. In this experiment, this parameter was set to 25. Since TiSASRec, DGCF, and DHCN are three SOTA methods that performed well on the three datasets in our experiment, we also compared PTGCN with the three baselines in the convergence rate.

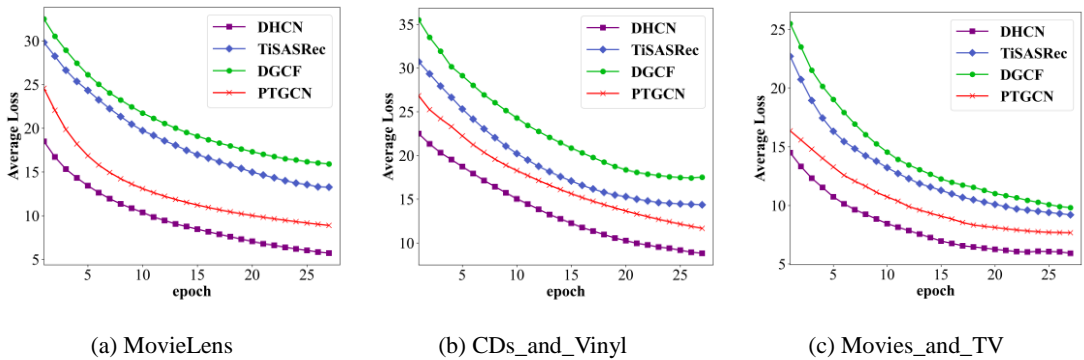

(a) MovieLens  (b) CDs_and_Vinyl  (c) Movies_and_TV

Fig. 7. Convergence analysis.

Table 7. Comparison among different methods in training efficiency.

| Dataset | Metric | GRU4Rec+ | Caser | SASRec | BERT4Rec | DHCN | TiSASRec | DGCF | PTGCN |
|---|---|---|---|---|---|---|---|---|---|
| | #Params | 2,762,426 | 6,105,060 | <u>911,840</u> | 1,724,861 | **727,680** | 950,560 | 39,147,668 | 1,842,880 |
| MovieLens | Time(ms)/batch | **13.9** | 143.1 | 25.4 | 31.6 | 706.3 | 34.4 | <u>16.5</u> | 22.1 |
| | $R@10$ | 0.6345 | 0.7064 | 0.7124 | 0.7215 | 0.7348 | 0.7591 | <u>0.7912</u> | **0.8746** |
| | $NDGG@10$ | 0.4252 | 0.4813 | 0.4871 | 0.4974 | 0.5012 | 0.5235 | <u>0.5673</u> | **0.6351** |
| Amazon CDs_and_Vinyl | Time(ms)/batch | **11.6** | 125.7 | 23.6 | 29.3 | 689.5 | 32.8 | <u>15.4</u> | 21.5 |
| | $R@10$ | 0.6012 | 0.6348 | 0.6438 | 0.6542 | 0.6645 | 0.6824 | <u>0.6917</u> | **0.7686** |
| | $NDGG@10$ | 0.3762 | 0.4189 | 0.4127 | 0.4238 | 0.4213 | 0.4486 | <u>0.4923</u> | **0.5342** |
| Amazon Movies_and_TV | Time(ms)/batch | **12.3** | 133.6 | 24.7 | 31.5 | 678.4 | 33.2 | <u>15.9</u> | 21.7 |
| | $R@10$ | 0.5829 | 0.6281 | 0.6139 | 0.6239 | 0.6349 | 0.6482 | <u>0.7347</u> | **0.8012** |
| | $NDGG@10$ | 0.3689 | 0.3992 | 0.4071 | 0.4213 | 0.4118 | 0.4267 | <u>0.5236</u> | **0.5722** |

Fig. 7 presents the convergence trends of PTGCN, TiSASRec, DGCF, and DHCN concerning the average loss on the three datasets. In each plot of Fig. 7, the X-axis denotes the epoch number, and the Y-axis represents the average loss. It is evident from Fig. 7 that the training losses of the four approaches continue to decline, but they are not substantially increased after 20 epochs and tend to be stable. During the training process of PTGCN, the loss value generated in the later stage (> 20 epochs) was much smaller than that generated in the early stage, which results in good convergence of evaluation metrics. In particular, the PTGCN model quickly converged to a stable state on the MoiveLens and Amazon Movies_and_TV datasets.

**5.5.4 Impacts of Temporal and Positional Encodings**

It has been recognized that time information and position information are two critical factors affecting the performance of sequential recommendation models. To further analyze the impacts of these two types of information on the recommendation performance of our method, we designed three variants of PTGCN, namely Non-GCN, PGCN, and TGCN. Non-GCN denotes that the PTGCN model does not consider any temporal and positional information, PGCN removes the time information fed to the PTGCN model, and TGCN only considers time information while disregarding the position information of interactions. More specifically, the time embedding and position embedding can be easily removed or added in the embedding layer of PTGCN. We then conducted an ablation study to analyze the difference between time embedding and position embedding through a detailed comparison among PTGCN, PGCN, and TGCN. Table 8 shows the recommendation performance of PTGCN and the three variants on the three datasets.

Table 8. Impacts of temporal and positional encodings on recommendation performance.

| Dataset | Model | $R@5$ | $NDGG@5$ | $R@10$ | $NDGG@10$ |
|---|---|---|---|---|---|
| MovieLens | Non-GCN | 0.7053 | 0.5351 | 0.8317 | 0.5763 |
| | PGCN | 0.7398 | 0.5721 | 0.8553 | 0.6098 |
| | TGCN | 0.7332 | 0.5612 | 0.8524 | 0.6001 |
| | PTGCN | **0.7651** | **0.5994** | **0.8746** | **0.6351** |
| Amazon CDs_and_Vinyl | Non-GCN | 0.5984 | 0.4436 | 0.7187 | 0.4517 |
| | PGCN | 0.6243 | 0.4678 | 0.7452 | 0.4786 |
| | TGCN | 0.6214 | 0.4637 | 0.7413 | 0.5102 |
| | PTGCN | **0.6421** | **0.4946** | **0.7686** | **0.5342** |
| Amazon Movies_and_TV | Non-GCN | 0. 6145 | 0.4878 | 0.7435 | 0.5129 |
| | PGCN | 0.6467 | 0.5154 | 0.7789 | 0.5478 |
| | TGCN | 0.6437 | 0.5125 | 0.7756 | 0.5436 |
| | PTGCN | **0.6784** | **0.5341** | **0.8012** | **0.5722** |

From Table 8, we have the following main findings:
- After leveraging time information or position information solely, our model's performance was consistently improved on the three datasets. Compared with Non-GCN, the four metrics of PGCN were, on average, increased by 4.82%, 6.01%, 3.76%, and 6.19%, respectively. Compared with Non-GCN, the four metrics of TGCN were, on average, increased by 4.18%, 4.82%, 3.32%, and 7.69%, respectively. Such results indicate that the time and position embeddings are helpful to improve the recommendation performance of our model.
- The performance of our model was also consistently improved on the three datasets when considering time information and position information simultaneously. Compared with PGCN, the four metrics of PTGCN were, on average, increased by 3.72%, 4.71%, 2.75%, and 6.74%, respectively. Compared with TGCN, PTGCN's metrics were, on average, increased by 4.36%, 5.90%, 3.20%, and 5.27%. Such results suggest that the position embedding does not overlap the time embedding in enhancing model performance. In other words, the impacts of these two types of information on our method are not equivalent, and they can be used together.

**5.5.5 Limitations**

As mentioned in Subsection 5.2, we filtered out a few inactive users and items in the three datasets to alleviate the cold-start problem, one of the most common problems in the field of recommender systems. Recently, Qian *et al.* [63] proposed a strict cold-start scenario, where some users/items do not appear in the training set and, at the same time, they do not have any user-item interactions during testing. The objective of this study is not to address the cold-start problem of new users or items. Considering the challenge of the sequential recommendation task, we did not test the recommendation performance of PTGCN in this extreme scenario. We need to improve PTGCN further to work in the strict cold-start recommendation scenario. Besides, PTGCN does not consider whether the recommended item is new for the target user. In other words, the recommended item is a new one with which the target user has never interacted. Although the next new item recommendation is helpful in some specific application scenarios [64], it is more challenging, and we will investigate it systematically.

The likely bias in user ratings or reviews exists in most publicly available datasets used for recommender systems, affecting recommendation approaches' performance to some extent. Unlike some recent studies [65], [66], we did not denoise such biased ratings or reviews to improve recommendation performance in this study. Besides, we have demonstrated the effectiveness of PTGCN on three real-world datasets of different sizes. In the three datasets, users vary from six thousand to 84 thousand. However, the scalability of PTGCN up to large-scale datasets containing millions of users remains unexplored. Therefore, one of our future works is to improve PTGCN from the above two aspects.

Performance evaluation using sampled metrics is a common practice for item recommendation. However, a recent work [67] shows that a few commonly-used sampled metrics, such as NDCG, Recall, and Average Precision, are high-bias and low-variance estimators to measure the performance of recommender systems, although they can speed up calculating metric values in research papers. In [67], Krichene and Rendle also proposed corrected metrics to improve the quality of the rank estimate, however, at the cost of increased variance. Considering that such corrected metrics have not been widely recognized in the academy and industry, we did not evaluate the recommendation performance of the proposed and baseline approaches by the corrected metrics in this work. However, we acknowledge that the results may become inconsistent with the exact metrics if we do not use the corrected metrics.

# 6 Conclusion

The sequential recommendation task is challenging for recommender systems to recommend the next item accurately. Although existing methods based on Markov chains and the RNN architecture can obtain good results, they have different shortcomings. We propose a new sequential recommendation model by introducing the GCN architecture in this article. In particular, the proposed model can simultaneously model the temporal dynamics of both users and items. Also, it can capture high-level interaction information (i.e., the higher-order connectivity) between users and items to generate more expressive representations for users and items. Empirical studies on three real-world datasets demonstrate the advantages of our model over competitive baselines in the trade-off between recommendation performance and training efficiency. Moreover, an ablation study validates the effectiveness and rationale of modeling the temporal dynamics of items and the high-order connectivity.

Our future work focuses on the following three aspects. First, self-supervised learning has great potential to enhance sequential recommendation quality [42], [43]. We plan to integrate self-supervised learning, such as SimCLR [68] and MoCo [69], into PTGCN and leverage more auxiliary information sources, such as knowledge graphs and social networks [70], in addition to time and position information. Second, the online session-based recommendation is challenging for recommender systems developed offline. We will explore more efficient solutions enabling PTGCN to work in online session-based scenarios for massive users. Third, since the strict cold-start recommendation is an interesting problem, we intend to improve PTGCN further so that it can work in such a scenario.

## Acknowledgment

This work was partially supported by the National Key Research and Development Program of China (No. 2020AAA0107705) and the National Science Foundation of China (Nos. 61972292 and 62006023). Yutao Ma is the corresponding author of this article.